\documentclass[
    ,final            
    ,numberedheadings 
  ]
  {aipproc}

\layoutstyle{8x11single}

\begin{document}

\title{Relative information entropy of an inhomogeneous universe}


\author{Masaaki Morita}{
  address={Department of Material Science and Technology, 
           Nagaoka University of Technology, 
           1603-1 Kamitomioka, Nagaoka, Niigata 940-2188, Japan}
}

\author{Thomas Buchert}{
  address={Universit\'e Lyon 1, Centre de Recherche Astrophysique de Lyon,
           CNRS UMR 5574, 9 avenue Charles Andr\'e, F-69230 
           Saint-Genis-Laval, France}
}

\author{Akio Hosoya}{
  address={Department of Physics, Tokyo Institute of Technology, 
           Oh-Okayama, Meguro, Tokyo 152-8551, Japan}
}

\author{Nan Li}{
  address={Center for Particle Physics and Phenomenology,
           Universit\'e catholique de Louvain, 
           2 Chemin du Cyclotron, B-1348 Louvain-la-Neuve, Belgium}
}

\begin{abstract}
In the context of averaging an inhomogeneous cosmological model, 
we propose a natural measure identical to the Kullback-Leibler 
relative information entropy, which expresses the distinguishability 
of the local inhomogeneous density field from its spatial average 
on arbitrary compact domains. 
This measure is expected to be an increasing function in time and 
thus to play a significant role in studying gravitational entropy. 
To verify this conjecture, we explore the time evolution of the measure 
using the linear perturbation theory of a spatially flat FLRW model 
and a spherically symmetric nonlinear solution. 
We discuss the generality and conditions for the time-increasing nature 
of the measure, and also the connection to the backreaction effect 
caused by inhomogeneities.
\end{abstract}

\maketitle


\section{Introduction}

Modern cosmology is based on the hypothesis 
called the Cosmological Principle, and the universe is assumed 
to be successfully described by a homogeneous and isotropic 
Friedmann-Lema\^itre-Robertson-Walker (FLRW) universe model 
on large scales. 
In spite of its simplicity, this hypothesis is highly non-trivial 
because a realistic universe model should include local inhomogeneities, 
and the physical property of such a realistic model 
averaged over a sufficiently large scale does not necessarily 
coincide with that of the FLRW universe. 
The difference between a spatially averaged inhomogeneous universe 
and the FLRW universe has been emphasized by Ellis~\cite{ellis1984}
and studied since the pioneering works 
by Futamase~\cite{futamase1988,futamase1996}. 
This topic is now widely noticed in the context of dark energy cosmology; 
the effect of inhomogeneities may be an alternative to introducing 
an exotic matter for the cosmic acceleration. 
(See, e.g. Ref.~\cite{kolb2006}; 
Refs.~\cite{rasanen2006,buchert2008} for comprehensive reviews.) 

In quantifying how a realistic inhomogeneous universe model 
departs from the FLRW one, 
it would be convenient to utilize some measure of inhomogeneity. 
In Information Theory, if we have two probability distributions, 
$\{ p_{i} \}$ and $\{ q_{i} \}$, and would like to quantify 
the \textit{distinguishability} of the two distributions, 
the relevant quantity is known to be the relative information entropy 
(sometimes called the \textit{Kullback-Leibler divergence})~\cite{ct1991}: 
\begin{equation}\label{KLentropy}
\mathcal{S}_{\mathrm{KL}} \{ p \parallel q \} := 
\sum_{i} p_{i} \ln \frac{p_{i}}{q_{i}} \, ,
\end{equation}
where $\{ p_{i} \}$ and $\{ q_{i} \}$ are the actual and presumed 
probability distributions, respectively. 
This relative entropy is positive for $q_{i} \neq p_{i}$, and zero 
if the two distributions $\{ p_{i} \}$ and $\{ q_{i} \}$ agree. 
Note that the $\mathcal{S}_{\mathrm{KL}} \{ p \parallel q \}$ is not 
symmetric for $\{ p_{i} \}$ and $\{ q_{i} \}$. 
With the help of the concept of the Kullback-Leibler relative 
information entropy, we proposed in our previous work~\cite{hbm2004} 
a natural measure of inhomogeneity in the universe, in the form 
\begin{equation}\label{RIE1}
\mathcal{S} \{ \varrho \parallel \langle \varrho \rangle_{\mathcal{D}} \} 
:= \int_{\mathcal{D}} \varrho 
\ln \frac{\varrho}{\langle \varrho \rangle_{\mathcal{D}}} 
\  \mathrm{d} \mu \, ,
\end{equation}
where $\varrho$ and $\langle \varrho \rangle_{\mathcal{D}}$ are 
respectively the actual matter distribution and its spatial average, 
$\mathrm{d} \mu$ is the Riemannian volume element, 
and the integration is performed on a compact spatial domain $\mathcal{D}$. 
Here we have adopted the formulation of averaging inhomogeneous universes 
developed by one of the authors~\cite{buchert2000}.
We also conjectured that the measure~(\ref{RIE1}) is an increasing 
function of time for sufficiently large times. 
Intuitively this measure is likely to have the time-increasing nature, 
but detailed analyses are required to show it. 
It is actually significant whether this conjecture is correct 
because the positive conclusion for the conjecture will lead to 
the validity of our measure being regarded as entropy 
in a gravitational system. 
The possibility that our measure is concerned 
with \textit{gravitational entropy} 
has been discussed in Ref.~\cite{elbu2005}. 

In this paper, we explore the time evolution of the measure 
using specific models of an inhomogeneous universe, 
and examine on what condition the time-increasing nature 
of the measure holds. 
We employ the linear perturbation theory of a spatially flat FLRW universe 
and a spherically symmetric nonlinear solution
as inhomogeneous universe models, and illustrate the temporal behavior 
of the measure explicitly with the models. 

This paper is organized as follows. In the next section, 
we give a brief review of the formalism of averaging inhomogeneous 
universes by following Ref.~\cite{buchert2000}. 
In the section~\ref{sec:RIE}, we introduce a measure of inhomogeneity 
analogous to the Kullback-Leibler divergence, 
and derive the time derivatives of the measure 
with a general consideration for the time-increasing nature of the measure. 
We present illustrative examples of the time evolution of the measure 
using the linear perturbation theory and a spherically symmetric exact 
solution, in the sections~\ref{sec:linear} and \ref{sec:LTB}, respectively. 
Finally in the section~\ref{sec:sum}, we summarize our results and 
give an outlook.

\section{Basics of averaging inhomogeneous universes}
\label{sec:basics}

Let us start by recalling the basic equations that govern the dynamics 
of a spatially averaged inhomogeneous universe, 
along the formulation developed 
by one of the authors~\cite{buchert2000,buchert2001}. 
We shall restrict our consideration, for simplicity, 
to an irrotational pressureless fluid with energy density $\varrho$ 
and four-velocity $u^{\mu}$, 
and work in a time-orthogonal foliation with the line element 
\begin{equation}
\mathrm{d} s^{2} = - \mathrm{d} t^{2} 
+ g_{ij} \mathrm{d} X^{i} \mathrm{d} X^{j} \, ,
\end{equation}
where $X^{i}$ are coordinates in the $t= \mathrm{const.}$ hypersurfaces 
(with three-metric $g_{ij}$) that are comoving with the fluid so that 
the four-velocity $u^{\mu} = (1, \mathbf{0})$. 
It is convenient for a description of the dynamics 
to use the expansion tensor $\Theta_{ij} := (1/2) \dot{g_{ij}}$, 
where an overdot $(\dot{\ })$ denotes time derivative, 
and its trace $\theta := g^{ij} \Theta_{ij}$ 
(the local expansion rate), and the traceless part 
$\sigma_{ij} := \Theta_{ij} - (1/3) \theta g_{ij}$ (the shear tensor). 
Using these quantities as dynamical variables, the continuity equation
and the Raychaudhuri equation are written as 
\begin{equation}\label{continuity-eq}
\dot{\varrho} + \varrho \theta = 0 \, ,
\end{equation}
\begin{equation}\label{raychaudhuri-eq}
\dot{\theta} = -4 \pi G \varrho - \frac{1}{3} \theta^{2} 
- 2 \sigma^{2} \, ,
\end{equation}
where $\sigma^{2} := (1/2) \sigma^{i}_{\ j} \sigma^{j}_{\ i}$ 
is the rate of shear squared.

We define averaging of a scalar quantity $A(t, X^{i})$ 
by the Riemannian volume average over a compact spatial domain $\mathcal{D}$:
\begin{equation}\label{average}
\langle A(t, X^{i}) \rangle_{\mathcal{D}}: = 
\frac{1}{V_{\mathcal{D}}} 
\int_{\mathcal{D}} A(t, X^{i}) \sqrt{g} \  \mathrm{d}^{3} X \  ; 
\qquad 
V_{\mathcal{D}}(t) : = \int_{\mathcal{D}} \sqrt{g} \  \mathrm{d}^{3} X \, ,
\end{equation}
with the Riemannian volume element 
$\mathrm{d} \mu:= \sqrt{g} \  \mathrm{d}^{3} X$, \  $g:=\det(g_{ij})$,
of the spatial hypersurfaces of constant time.
We also introduce an effective scale factor via the volume 
(normalized by the volume of the initial domain 
$V_{{\mathcal{D}}_{\mathrm{i}}}$), 
$a_{\mathcal{D}}(t) : = 
( V_{\mathcal{D}}(t) / V_{{\mathcal{D}}_{\mathrm{i}}} )^{1/3}$.
Then the averaged expansion rate is expressed in terms of the effective 
scale factor as
\begin{equation}
\langle \theta \rangle_{\mathcal{D}} = 
\frac{\dot{V}_{\mathcal{D}}}{V_{\mathcal{D}}} = 
3 \frac{\dot{a}_{\mathcal{D}}}{a_{\mathcal{D}}} \, .
\end{equation}
The key concept in the averaging formalism is \emph{non-commutativity} 
of two operations, spatial average and time evolution. 
This is expressed by a \textit{commutation rule} for the averaging 
of a scalar field $A$~\cite{buchert2000,buchert2001,bueh1997,rskb1997}: 
\begin{equation}\label{commutation}
\frac{\partial}{\partial t} \langle A \rangle_{\mathcal{D}} - 
\left\langle \frac{\partial A}{\partial t} \right\rangle_{\mathcal{D}}
= \langle A \theta \rangle_{\mathcal{D}} 
- \langle A \rangle_{\mathcal{D}} \langle \theta \rangle_{\mathcal{D}}
= \langle \delta A \, \delta \theta \rangle_{\mathcal{D}} \, ,
\end{equation}
where $\delta A := A - \langle A \rangle_{\mathcal{D}}$ and 
$\delta \theta := \theta - \langle \theta \rangle_{\mathcal{D}}$ 
represent the deviations of local values of the fields from their averages.
Averaging Eqs.~(\ref{continuity-eq}) and (\ref{raychaudhuri-eq}) 
with the help of Eq.~(\ref{commutation}) yields 
\begin{equation}\label{ave-continuity}
\frac{\partial}{\partial t} \langle \varrho \rangle_{\mathcal{D}} 
+ \langle \varrho \rangle_{\mathcal{D}} \langle \theta \rangle_{\mathcal{D}}
= 0 \, ,
\end{equation}
\begin{equation}\label{ave-raychaudhuri}
3 \frac{\ddot{a}_{\mathcal{D}}}{a_{\mathcal{D}}} 
+ 4 \pi G \langle \varrho \rangle_{\mathcal{D}} = Q_{\mathcal{D}} \  ; 
\qquad Q_{\mathcal{D}} := 
\frac{2}{3} \left( \langle \theta^{2} \rangle_{\mathcal{D}} 
- \langle \theta \rangle_{\mathcal{D}}^{2} \right)
- 2 \langle \sigma^{2} \rangle_{\mathcal{D}} \, ,
\end{equation}
where $Q_{\mathcal{D}}$ is the `kinematical backreaction term', 
which appears due to inhomogeneities of cosmic matter distribution 
and leads the effective cosmic expansion given by $a_{\mathcal{D}}$ 
to deviate from the Friedmannian one. 
Equation~(\ref{ave-raychaudhuri}) tells us that the kinematical 
backreaction term consists of the fluctuation of the expansion rate 
and the averaged shear rate; 
the former plays the role of effective negative pressure, 
and the latter can be regarded as additional matter density.

\section{Relative information entropy}
\label{sec:RIE}

In order to introduce a quantity that measures how the universe is 
inhomogeneous within the formulation explained in the previous section, 
we pay particular attention to the commutation rule 
for the matter density field: 
\begin{equation}\label{commutation-rho}
\frac{\partial}{\partial t} \langle \varrho \rangle_{\mathcal{D}} - 
\left\langle \frac{\partial \varrho}{\partial t} \right\rangle_{\mathcal{D}}
= \langle \varrho \theta \rangle_{\mathcal{D}} 
- \langle \varrho \rangle_{\mathcal{D}} \langle \theta \rangle_{\mathcal{D}}
= \langle \delta \varrho \, \delta \theta \rangle_{\mathcal{D}} \, .
\end{equation}
This means that the time evolution of the averaged density field 
does not coincide with the average of the density field evolved locally. 
We consider that the difference between 
$\langle \varrho \rangle_{\mathcal{D}}^{\cdot}$ 
and $\langle \dot{\varrho} \rangle_{\mathcal{D}}$ leads to the entropy 
production for the matter density field. 
This idea brings us to write 
\begin{equation}\label{commutation-dotS}
\frac{\partial}{\partial t} \langle \varrho \rangle_{\mathcal{D}} - 
\left\langle \frac{\partial \varrho}{\partial t} \right\rangle_{\mathcal{D}}
= - \frac{\dot{\mathcal{S}}}{V_{\mathcal{D}}} \, ,
\end{equation}
where $\mathcal{S}$ is an entropy associated with the density field. 
Looking for a functional of the matter density field that satisfies 
Eq.~(\ref{commutation-dotS}), we find that, 
interestingly, the answer is~\cite{hbm2004}: 
\begin{equation}
\mathcal{S} \{ \varrho \parallel \langle \varrho \rangle_{\mathcal{D}} \} 
:= \int_{\mathcal{D}} \varrho 
\ln \frac{\varrho}{\langle \varrho \rangle_{\mathcal{D}}} \sqrt{g} 
\  \mathrm{d}^{3} X \, ,
\end{equation}
which is analogous to the Kullback-Leibler relative information entropy, 
Eq.~(\ref{KLentropy}). 
Note that, for strictly positive density, $\varrho > 0$, the entropy 
$\mathcal{S} \{ \varrho \parallel \langle \varrho \rangle_{\mathcal{D}} \}$ 
is positive definite if $\varrho \neq \langle \varrho \rangle_{\mathcal{D}}$, 
and $\mathcal{S} = 0$ if and only if 
$\varrho = \langle \varrho \rangle_{\mathcal{D}}$.

Let us explore the temporal behavior of the relative information entropy 
$\mathcal{S} \{ \varrho \parallel \langle \varrho \rangle_{\mathcal{D}} \}$ 
to verify whether the $\mathcal{S}$ possesses the time-increasing nature. 
 From Eqs.~(\ref{commutation-rho}) and (\ref{commutation-dotS}), 
the time derivative of the entropy is immediately found to give 
\begin{equation}\label{dotS}
\frac{\mathrm{d}}{\mathrm{d} t} 
\mathcal{S} \{ \varrho \parallel \langle \varrho \rangle_{\mathcal{D}} \}
= - \int_{\mathcal{D}} \delta \varrho \, \delta \theta 
\sqrt{g} \  \mathrm{d}^{3} X 
= - V_{\mathcal{D}} 
\langle \delta \varrho \, \delta \theta \rangle_{\mathcal{D}} \, .
\end{equation}
We expect from Eq.~(\ref{dotS}) that the time derivative of $\mathcal{S}$ 
will generally be positive in view of cosmological structure formation, 
because, on average, an overdense region ($\delta \varrho > 0$) tends to 
contract ($\delta \theta < 0$) to form a cluster, 
and an underdense region ($\delta \varrho < 0$) tends to expand 
($\delta \theta > 0$) to form a void. 
To be more precise, however, how inhomogeneities evolve depends on 
initial conditions, particularly at an early stage of the evolution. 
 For example, a fluid element included in an overdense region does not 
necessarily have an initial expansion rate with the direction of contracting, 
because the density field and the local expansion rate are given 
independently in the initial data setting. 
At a sufficiently late stage, the effect of initial conditions will 
get weaker and inhomogeneities will evolve according to the intuitive manner 
as we mentioned above, leading to the positivity of the time derivative 
of the entropy. It is therefore plausible that, even if the time derivative 
of the entropy is negative temporarily, it will become positive eventually. 
This idea implies the importance of examining 
whether the second time derivative of the entropy is positive, 
i.e. the \textit{time-convexity} of the entropy. 
Differentiation of Eq.~(\ref{dotS}), together with 
Eqs.~(\ref{raychaudhuri-eq}) and (\ref{commutation}), yields 
\begin{equation}\label{ddotS1}
\frac{1}{V_{\mathcal{D}}} \frac{\mathrm{d}^{2}}{\mathrm{d} t^{2}} 
\mathcal{S} \{ \varrho \parallel \langle \varrho \rangle_{\mathcal{D}} \}
= 4 \pi G \left( \Delta \varrho \right)^{2} 
+ \langle \varrho \rangle_{\mathcal{D}} \left( \Delta \theta \right)^{2}
+ \frac{1}{3} \langle \delta \varrho \, \delta \theta^{2} \rangle_{\mathcal{D}}
+ 2 \langle \delta \varrho \, \delta \sigma^{2} \rangle_{\mathcal{D}} \, ,
\end{equation}
where $\Delta \varrho := 
\sqrt{\langle (\delta \varrho)^{2} \rangle_{\mathcal{D}}}$ and 
$\Delta \theta := 
\sqrt{\langle (\delta \theta)^{2} \rangle_{\mathcal{D}}}$ 
are the fluctuation amplitudes of density and expansion, respectively. 
Using the formula 
\begin{equation}
\delta (\theta^{2}) =
2 \langle \theta \rangle_{\mathcal{D}} \delta \theta 
+ (\delta \theta)^{2} - (\Delta \theta)^{2} \, ,
\end{equation}
the second time derivative of the entropy, Eq.~(\ref{ddotS1}), 
is rewritten as 
\begin{eqnarray}
\frac{\ddot{\mathcal{S}}}{V_{\mathcal{D}}}
&=& 4 \pi G \left( \Delta \varrho \right)^{2} 
+ \frac{1}{3} \langle \varrho (\delta \theta)^{2} \rangle_{\mathcal{D}}
+ \frac{2}{3} \langle \varrho \rangle_{\mathcal{D}} 
  \left( \Delta \theta \right)^{2}
+ 2 \langle \delta \varrho \, \delta \sigma^{2} \rangle_{\mathcal{D}} 
+ \frac{2}{3} \langle \theta \rangle_{\mathcal{D}}
\langle \delta \varrho \, \delta \theta \rangle_{\mathcal{D}}
\label{ddotS2} \\
&=& 4 \pi G \left( \Delta \varrho \right)^{2} 
+ \frac{1}{3} \langle \varrho (\delta \theta)^2 \rangle_{\mathcal{D}}
+ 2 \langle \varrho \sigma^{2} \rangle_{\mathcal{D}}
+ \langle \varrho \rangle_{\mathcal{D}} Q_{\mathcal{D}}
- \frac{2}{3} \langle \theta \rangle_{\mathcal{D}}
\frac{\dot{\mathcal{S}}}{V_{\mathcal{D}}} \, .
\label{ddotS3}
\end{eqnarray}

In order to clarify the conditions under which the positivity of 
$\dot{\mathcal{S}}$ holds, 
the sign of the second time derivative $\ddot{\mathcal{S}}$ is crucial, 
in particular at the instant $t = t_{\mathrm{c}}$ 
when $\dot{\mathcal{S}} = 0$. 
If $\ddot{\mathcal{S}} (t = t_{\mathrm{c}})$ is shown to be positive, 
we can conclude that $\dot{\mathcal{S}}$ is always positive thereafter. 
Note that, from Eq.~(\ref{ddotS3}), this applies to the case 
when the backreaction term 
$Q_{\mathcal{D}}$ is positive at $t = t_{\mathrm{c}}$.
One typical behavior of $\dot{\mathcal{S}}$ will be the following: 
suppose that $\dot{\mathcal{S}}$ is negative at some time 
for an averaging domain $\mathcal{D}$ large enough for 
the averaged expansion rate $\langle \theta \rangle_{\mathcal{D}} > 0$ 
and the averaged shear rate 
$\langle \sigma^{2} \rangle_{\mathcal{D}} \approx 0$. 
Then from Eq.~(\ref{ddotS3}), $\ddot{\mathcal{S}}$ is positive 
and thus $\dot{\mathcal{S}}$ increases with time. 
If $\dot{\mathcal{S}}$ is not bounded by a negative value, 
$\dot{\mathcal{S}}$ reaches zero at $t = t_{\mathrm{c}}$
and stays positive thereafter.

\section{Evolution of the entropy in the linear regime}
\label{sec:linear}

In what follows, we illustrate the temporal behavior of the entropy 
$\mathcal{S}$ by using specific models. 
 First let us observe the positivity of $\dot{\mathcal{S}}$ 
using the linear perturbation theory, 
which is valid in the early stage of structure formation in the universe, 
where the deviation from a homogenous and isotropic background is small. 
This illustrative calculation with the linear perturbation theory is, 
in a sense, most intuitive and easiest to carry out. 
The average properties of an inhomogeneous universe constructed 
with the linear perturbation theory have been investigated 
in Ref.~\cite{lischwarz2007}. 
We follow Ref.~\cite{lischwarz2007} in the method of the calculation 
performed in this section. 

Suppose that the three-metric is written as 
$g_{ij} = a(t)^2 (\gamma_{ij} + h_{ij})$ 
with the Friedmannian scale factor $a(t)$, 
the background metric $\gamma_{ij}$, 
and metric perturbation $h_{ij}$, 
and a scalar quantity $A(t, X^{i})$ is divided into 
a homogeneous part $A_{\mathrm{H}}(t)$ and an inhomogeneous perturbation 
$A_{\mathrm{p}}(t, X^{i})$, which is associated with $h_{ij}$.
The spatial average $\langle A \rangle_{\mathcal{D}}$ is then 
represented in terms of perturbation variables as
\begin{equation}
\langle A \rangle_{\mathcal{D}} := \frac{1}{V_{\mathcal{D}}} 
\int_{\mathcal{D}} A \sqrt{g} \  \mathrm{d}^{3} X 
= A_{\mathrm{H}} + \langle A_{\mathrm{p}} \rangle_{\gamma} 
  + \frac{1}{2} \left( \langle A_{\mathrm{p}} h \rangle_{\gamma}
  - \langle A_{\mathrm{p}} \rangle_{\gamma} \langle h \rangle_{\gamma} 
  \right) + \mathcal{O}(h^{3}) \, ,
\end{equation}
where $\langle \cdot \rangle_{\gamma} := 
\int_{\mathcal{D}} (\cdot) \sqrt{\gamma} \  \mathrm{d}^{3} X
/ \int_{\mathcal{D}} \sqrt{\gamma} \  \mathrm{d}^{3} X$ 
($\gamma := \det(\gamma_{ij})$) denotes 
the average defined on the background three-space, 
and $h := \gamma^{ij} h_{ij}$. 
Introducing another scalar quantity 
$B(t, X^{i}) = B_{\mathrm{H}}(t) + B_{\mathrm{p}}(t, X^{i})$ 
with a homogeneous part $B_{\mathrm{H}}(t)$ and 
a perturbation $B_{\mathrm{p}}(t, X^{i})$, 
we have the following useful formula:
\begin{equation}\label{delA-delB}
\langle \delta A \, \delta B \rangle_{\mathcal{D}} = 
\langle A_{\mathrm{p}} B_{\mathrm{p}} \rangle_{\gamma}
- \langle A_{\mathrm{p}} \rangle_{\gamma} 
  \langle B_{\mathrm{p}} \rangle_{\gamma} + \mathcal{O}(h^{3}) \, .
\end{equation}

The scalar-mode solution for the linear metric perturbation $h_{ij}$ 
in a spatially flat background without a cosmological constant 
is~\cite{lischwarz2007,peebles1980} 
\begin{equation}\label{linearmetric}
h_{ij} = \frac{20}{9} \Psi \gamma_{ij} + 2 D_{+}(t) \Psi_{|ij}
+ 2 D_{-}(t) \Phi_{|ij} \, ,
\end{equation}
where $\Psi = \Psi(X^{i})$ and $\Phi = \Phi(X^{i})$ are arbitrary 
functions of only spatial coordinates $X^{i}$, 
determined by initial conditions, 
\ $|$ denotes covariant derivative with respect to $\gamma_{ij}$, 
and $D_{+}(t) = t^{2/3}$ and $D_{-}(t) = t^{-1}$ are time-dependent 
factors of the growing and decaying modes, respectively. 
The energy density and the expansion rate are then given as 
\begin{eqnarray}
\label{linear-rho}
\varrho &=& \varrho_{\mathrm{H}}
\left(1 - t^{2/3} \nabla^{2} \Psi - t^{-1} \nabla^{2} \Phi
\right) \, , \\
\theta &=& 3 \frac{\dot{a}}{a}
+ \frac{2}{3} t^{-1/3} \nabla^{2} \Psi - t^{-2} \nabla^{2} \Phi \, ,
\label{linear-theta}
\end{eqnarray}
where $\varrho_{\mathrm{H}} = \varrho_{\mathrm{H}}(t)$ is 
the energy density of the homogeneous background, 
and $\nabla^{2} (\cdot) := \gamma^{ij} (\cdot)_{|ij}$.

Inserting Eqs.~(\ref{linear-rho}) and (\ref{linear-theta}) 
into Eq.~(\ref{delA-delB}), 
the time derivative of the entropy is calculated straightforwardly as, 
to the leading order, 
\begin{eqnarray}
4 \pi G \frac{\dot{\mathcal{S}}}{V_{\mathcal{D}}}
&=& \frac{4}{9} t^{-5/3} \left[ \langle (\nabla^{2} \Psi)^{2} \rangle_{\gamma} 
 - \langle \nabla^{2} \Psi \rangle_{\gamma}^{2} \right] 
 - \frac{2}{9} t^{-10/3} 
 \left[ \langle \nabla^{2} \Psi \  \nabla^{2} \Phi \rangle_{\gamma} -
 \langle \nabla^{2} \Psi \rangle_{\gamma} 
 \langle \nabla^{2} \Phi \rangle_{\gamma} \right] \nonumber \\
&& - \frac{2}{3} t^{-5} \left[ \langle (\nabla^{2} \Phi)^{2} \rangle_{\gamma} 
 - \langle \nabla^{2} \Phi \rangle_{\gamma}^{2} \right] \, .
\label{dotSlinear}
\end{eqnarray}
Equation~(\ref{dotSlinear}) tells us that, 
if the dynamics of inhomogeneity is dominated by the growing mode, 
$\dot{\mathcal{S}}$ is positive, 
meaning that the universe becomes more and more inhomogeneous. 
At early times, the decaying mode may dominate the growing one, 
and then $\dot{\mathcal{S}} < 0$, which implies that 
the universe can temporarily become less and less inhomogeneous, 
but for large enough times, the growing mode is expected to dominate 
and $\dot{\mathcal{S}} > 0$ eventually. 
One exception is the case where there is no growing mode, 
but it is quite rare and unphysical, and we can ignore it safely.
Thus on the positivity of $\dot{\mathcal{S}}$ in the linear regime, 
we can claim that 
\emph{the time derivative $\dot{\mathcal{S}}$ becomes positive 
at least for sufficiently large times, unless the growing mode 
is exactly zero.}

In order to show the above fact from another point of view, 
let us consider the second time derivative $\ddot{\mathcal{S}}$, 
Eq.~(\ref{ddotS1}). 
Estimating each term of the right-hand side of Eq.~(\ref{ddotS1}) 
with the linear perturbation theory, 
we find that the leading order of the last term is $\mathcal{O}(h^{3})$ 
while that of the other terms is $\mathcal{O}(h^{2})$. 
Hence the last term can be neglected in this consideration. 
In addition, we can utilize the following approximate formula 
in the linear regime: 
\begin{equation}
\langle \delta \varrho \, \delta \theta^{2} \rangle_{\mathcal{D}}
\approx 2 \langle \theta \rangle_{\mathcal{D}}
\langle \delta \varrho \, \delta \theta \rangle_{\mathcal{D}} \, ,
\end{equation}
and thus we obtain 
\begin{equation}\label{ddotS-approx}
\frac{\ddot{\mathcal{S}}}{V_{\mathcal{D}}}
\approx 4 \pi G \left( \Delta \varrho \right)^{2} 
+ \langle \varrho \rangle_{\mathcal{D}} \left( \Delta \theta \right)^{2}
+ \frac{2}{3} \langle \theta \rangle_{\mathcal{D}}
\langle \delta \varrho \, \delta \theta \rangle_{\mathcal{D}} \, .
\end{equation}
We find from Eq.~(\ref{ddotS-approx}) that, 
when $\dot{\mathcal{S}} < 0$, the second derivative 
$\ddot{\mathcal{S}}$ is positive 
because all terms in the right-hand side of Eq.~(\ref{ddotS-approx}) 
are positive; 
if $\dot{\mathcal{S}} = 0$ at an instant $t=t_{\mathrm{c}}$, 
the second derivative $\ddot{\mathcal{S}}$ at $t=t_{\mathrm{c}}$ 
is positive. 
Therefore, $\dot{\mathcal{S}}$ is always going to be positive 
at $t=t_{\mathrm{c}}$, and thereafter stays positive, 
as far as the dynamics is in the linear regime. 

These facts are also understood directly by an explicit form of
the second time derivative, written in terms of the linear perturbations.
It is actually given as, to the leading order, 
\begin{eqnarray}
4 \pi G \frac{\ddot{\mathcal{S}}}{V_{\mathcal{D}}}
&=& \frac{4}{27} t^{-8/3} \left[ \langle (\nabla^{2} \Psi)^{2} \rangle_{\gamma}
 - \langle \nabla^{2} \Psi \rangle_{\gamma}^{2} \right] 
 + \frac{8}{27} t^{-13/3} \left[ \langle \nabla^{2} \Psi 
   \  \nabla^{2} \Phi \rangle_{\gamma} -
 \langle \nabla^{2} \Psi \rangle_{\gamma} 
 \langle \nabla^{2} \Phi \rangle_{\gamma} \right] \nonumber \\
&& + 2 t^{-6} \left[ \langle (\nabla^{2} \Phi)^{2} \rangle_{\gamma} 
 - \langle \nabla^{2} \Phi \rangle_{\gamma}^{2} \right] \\
&=& \frac{4}{27} t^{-8/3}
\left[ \left\langle (\nabla^{2} \Psi + t^{-5/3} \nabla^{2} \Phi)^{2} 
 \right\rangle_{\gamma}
 - \left\langle \nabla^{2} \Psi + t^{-5/3} \nabla^{2} \Phi 
\right\rangle_{\gamma}^{2} \right] \nonumber \\
&& + \frac{50}{27} t^{-6} \left[ \langle (\nabla^{2} \Phi)^{2} 
 \rangle_{\gamma} - \langle \nabla^{2} \Phi \rangle_{\gamma}^{2} 
\right] \quad \geq 0 \, .
\end{eqnarray}
Hence we obtain the following proposition in the linear regime: 
\emph{The second time derivative $\ddot{\mathcal{S}}$ is 
positive definite in the linear regime. 
Therefore, even if $\dot{\mathcal{S}}$ is negative for a while, 
$\dot{\mathcal{S}}$ is always going to be positive. 
Once $\dot{\mathcal{S}}$ arrives at zero, 
$\dot{\mathcal{S}}$ is always positive thereafter during the linear regime.}

\section{Nonlinear example with the LTB model}
\label{sec:LTB}

We proceed in this section to explore the time evolution of 
the relative information entropy 
$\mathcal{S} \{\varrho \parallel \langle \varrho \rangle_{\mathcal{D}} \}$ 
in the nonlinear regime with the spherically symmetric 
Lema\^\i tre--Tolman--Bondi (LTB) solution. 
This exact solution of Einstein's equations is often employed as a model of 
an inhomogeneous universe~\cite{krasinski1997,enqvist2008,hellaby2009,celerier2010}. 
The line element of the LTB solution reads 
\begin{equation}
\mathrm{d} s^{2} = -\mathrm{d} t^{2} 
+ \frac{r^{\prime 2}}{1+f} \mathrm{d} R^{2}
+ r^{2} (\mathrm{d} \vartheta^{2} 
+ \sin^{2} \vartheta \mathrm{d} \phi^{2}) \,,
\end{equation}
where a prime $(\ ')$ denotes $\partial / \partial R$, 
$f=f(R)$ is an arbitrary function of the comoving radial coordinate $R$ 
with $f>-1$, and $r=r(t,R)$.
Then Einstein's equations yield 
\begin{equation}\label{rho-spherical}
8 \pi G \varrho = \frac{F'}{r' r^2} \, ,
\end{equation}
\begin{equation}\label{dotr2}
\dot{r}^2 = \frac{F(R)}{r} + f(R) \, ,
\end{equation}
where $F(R)$ is another arbitrary function, 
which represents the initial mass distribution. 
The solutions of Eq.~(\ref{dotr2}) can be expressed 
in the following form: 

(i) $f>0$:
\begin{equation}
r = \frac{F}{2f} (\cosh \eta -1) \, , \qquad
t - T(R) = \frac{F}{2f^{3/2}} (\sinh \eta - \eta) \, ,
\end{equation}

(ii) $f<0$:
\begin{equation}
r = \frac{F}{-2f} (1 - \cos \eta) \, , \qquad
t - T(R) = \frac{F}{2(-f)^{3/2}} (\eta - \sin \eta) \, ,
\end{equation}

(iii) $f=0$:
\begin{equation}
r = \left(\frac{9F}{4}\right)^{1/3}
    \left( t-T(R) \right)^{2/3} \, ,
\end{equation}
where $T(R)$ is a third arbitrary function 
that comes from integrating Eq.~(\ref{dotr2}). 
In any of these three cases, the expansion rate and 
the shear rate squared are given as
\begin{equation}\label{theta-spherical}
\theta = \frac{\dot{r}'}{r'} + 2 \frac{\dot r}{r} 
= \frac{(r^{2} r')^{\cdot}}{r^{2} r'} \, , \qquad
\sigma^{2} = \frac{1}{3} 
\left( \frac{\dot{r}'}{r'} - \frac{\dot{r}}{r} \right)^{2} \, .
\end{equation}
If we regard the LTB solution as a spatially flat FLRW universe 
plus spherical linear perturbations, and make a connection between 
the functions $f(R)$, $F(R)$, and $T(R)$ in the LTB solution, 
and the functions $\Psi$ and $\Phi$ that appear 
in Eq.~(\ref{linearmetric}) in the linear perturbation theory, 
we obtain~\cite{mnk1998}: 
\begin{equation}
f = \frac{20}{9} \Psi' R \, , \qquad
F = \frac{4}{9} R^{3} \left( 1 + \frac{10}{3} \Psi \right) \, , \qquad 
T = - \frac{3}{2} \Phi' R^{-1} \, .
\end{equation}
These relations tell us that the functions $f(R)$ and $T(R)$ 
correspond to the growing and decaying modes in the linear perturbation 
theory, respectively. 

Let us write averaged dynamical variables in the LTB solution. 
Taking a spherical compact domain ${\mathcal{D}}$: 
\[
{\mathcal{D}} = \left\{ (R, \, \vartheta, \, \phi) |
                 \ 0 \leq R \leq R_{\mathrm{d}} ,
                 \ 0 \leq \vartheta \leq \pi ,
                 \ 0 \leq \phi \leq 2\pi
                \right\} \, ,
\]
with the comoving radius of the spherical domain $R_{\mathrm{d}}$, 
we have
\begin{equation}\label{ave-rho-spherical}
8 \pi G \langle \varrho \rangle_{\mathcal{D}}
= \left\langle \frac{F'}{r^{2} r'} \right\rangle_{\mathcal{D}}
= \frac{4 \pi}{V_{\mathcal{D}}} 
\int_{0}^{R_{\mathrm{d}}} \frac{F'}{(1+f)^{1/2}} \  \mathrm{d} R \, ,
\end{equation}
\begin{equation}\label{ave-theta-spherical}
\langle \theta \rangle_{\mathcal{D}} = 
\left\langle \frac{\dot{r}'}{r'} + 2 \frac{\dot r}{r} 
\right\rangle_{\mathcal{D}}
= \frac{4 \pi}{V_{\mathcal{D}}} 
\int_{0}^{R_{\mathrm{d}}} \frac{(r^{2} r')^{\cdot}}{(1+f)^{1/2}} 
\  \mathrm{d} R \, ,
\end{equation}
with the volume of the domain $\mathcal{D}$, 
\begin{equation}
V_{\mathcal{D}} = 4 \pi 
\int_{0}^{R_{\mathrm{d}}} \frac{r^{2} r'}{(1+f)^{1/2}} 
\  \mathrm{d} R \, .
\end{equation}
Using Eqs.~(\ref{rho-spherical}), (\ref{theta-spherical}), 
(\ref{ave-rho-spherical}), and (\ref{ave-theta-spherical}), 
the backreaction term $Q_{\mathcal{D}}$ and the time derivative of 
the entropy in the spherical case become
\begin{equation}
Q_{\mathcal{D}} = \left\langle 2 \frac{\dot{r}}{r}
\left( 2 \frac{\dot{r}'}{r'} + \frac{\dot{r}}{r} \right)
\right\rangle_{\mathcal{D}}
- \left\langle \frac{\dot{r}'}{r'} + 2 \frac{\dot{r}}{r}
\right\rangle_{\mathcal{D}}^{2} \, ,
\end{equation}
\begin{equation}
8 \pi G \dot{\mathcal{S}} = 
\frac{4 \pi \dot{V}_{\mathcal{D}}}{V_{\mathcal{D}}} 
\int_{0}^{R_{\mathrm{d}}} \frac{F' \  \mathrm{d} R}{(1+f)^{1/2}}
- 4 \pi \int_{0}^{R_{\mathrm{d}}} 
\frac{F' (r^{2} r')^{\cdot} \  \mathrm{d} R}{r^{2} r' (1+f)^{1/2}} \, .
\end{equation}

In order to illustrate the time evolution of the entropy, 
we choose the three arbitrary functions, as is suggested 
in Ref.~\cite{mathum2001}, as 
\begin{equation}
T(R) = 0 \quad \mbox{for all } \  R \geq 0 \, , \qquad 
f(R) = \left\{ 
       \begin{array}{cl}
       -\left( \frac{R}{R_{0}} \right)^{2} 
        \left( 1 - \frac{R}{R_{0}} \right)^{2}
       & \quad \mbox{for } \  0 \leq R \leq R_{0} \, , \\
     0 & \quad \mbox{for } \  R > R_{0} \, ,
       \end{array}
       \right.
\end{equation}
\begin{equation}
F(R) = \left\{ 
       \begin{array}{ll}
       \frac{4}{9} R^{3}
       & \quad \mbox{for } \  0 \leq R < R_{1} \, , \\
       \frac{4}{9} R_{1}^{3}
       & \quad \mbox{for } \  R_{1} \leq R \leq R_{0} \, , \\
       \frac{4}{9} R^{3} \left( \frac{R_{1}}{R_{0}} \right)^{3}
       & \quad \mbox{for } \  R > R_{0} \, ,
       \end{array}
       \right.
\end{equation}
where $R_{0}$ and $R_{1}$ are constants that satisfy $R_{1} > R_{0} / 2$. 
The inner region ($0 \leq R \leq R_{1}$) of this model is 
the collapsing LTB solution, which is surrounded by a vacuum region 
($R_{1} < R \leq R_{0}$), and the outermost region ($R > R_{0}$) is 
the Einstein-de Sitter universe, 
i.e. a spatially flat FLRW universe without a cosmological constant. 
Setting $T(R) = 0$ is appropriate for our purpose, 
because the function $T(R)$ corresponds to the decaying mode 
in the linear perturbation theory, and its effect disappears 
 for sufficiently large times. 

\begin{figure}\label{adfig}
  \includegraphics[height=.30\textheight]{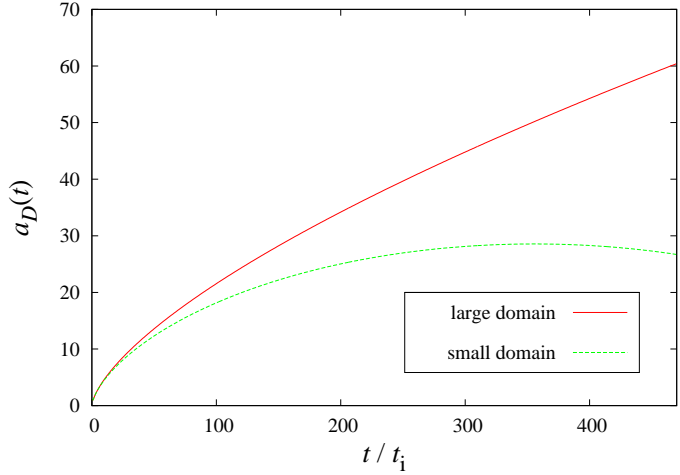}
  \caption{The effective scale factors for the spherical domains 
           are plotted as functions of time. 
           The solid line is for the large domain, and the dashed one 
           for the small domain.}
\end{figure}

In our illustration, we choose the constants $R_{0}$ and $R_{1}$ 
so that $R_{1} = 3 R_{0} / 4$, 
and consider two cases of the radius $R_{\mathrm{d}}$ of the averaging domain: 
(i) $R_{\mathrm{d}} = R_{0}$, which is called `the large domain', 
and (ii) $R_{\mathrm{d}} = R_{0} / 8$, `the small domain'. 
Note that, in the case of the large domain, the averaging domain contains 
the collapsing LTB region and the vacuum region, 
while the small domain consists of only the collapsing LTB region. 
As the result of the choice of the domain, the evolution of 
the effective scale factor for the large (small) domain is similar to 
that of a spatially flat (closed) FLRW universe, 
as we see in Fig~\ref{adfig}. 

\begin{figure}\label{brfig}
  \includegraphics[height=.30\textheight]{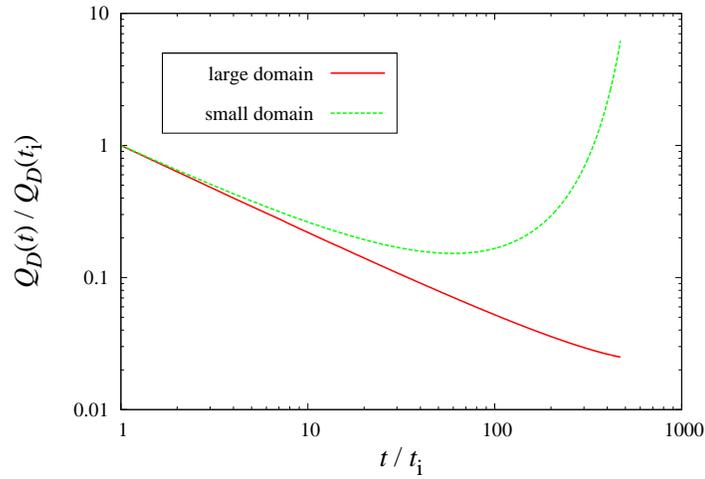}
  \caption{The time evolution of the backreaction term divided by 
           its initial value is plotted. 
           The solid line is for the large domain, 
           and the dashed one for the small domain.}
\end{figure}

We also present the time evolution of the kinematical backreaction 
term divided by its initial value for the large and small domains 
in Fig.~\ref{brfig}. 
In both cases, the backreaction term has, in fact, negative values 
through the evolution, 
and thus we find that the results shown in Fig.~\ref{brfig} are consistent 
with those given in Fig~\ref{adfig}. 

\begin{figure}\label{dotSlarge}
  \includegraphics[height=.30\textheight]{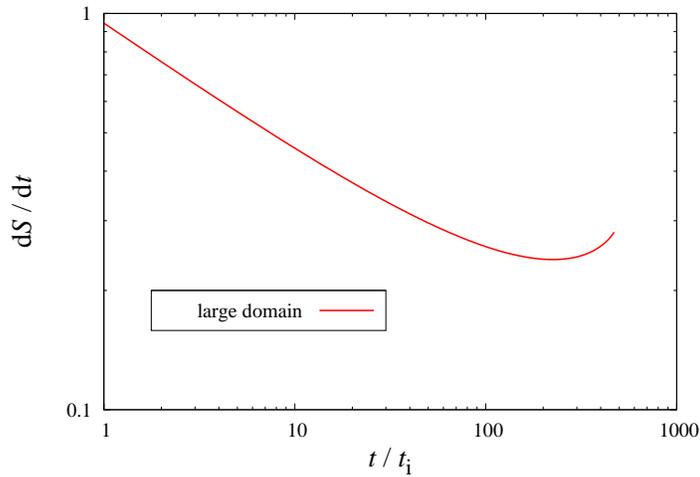}
  \caption{The time derivative of the entropy for the large domain 
           is plotted as a function of time.}
\end{figure}

\begin{figure}\label{dotSsmall}
  \includegraphics[height=.30\textheight]{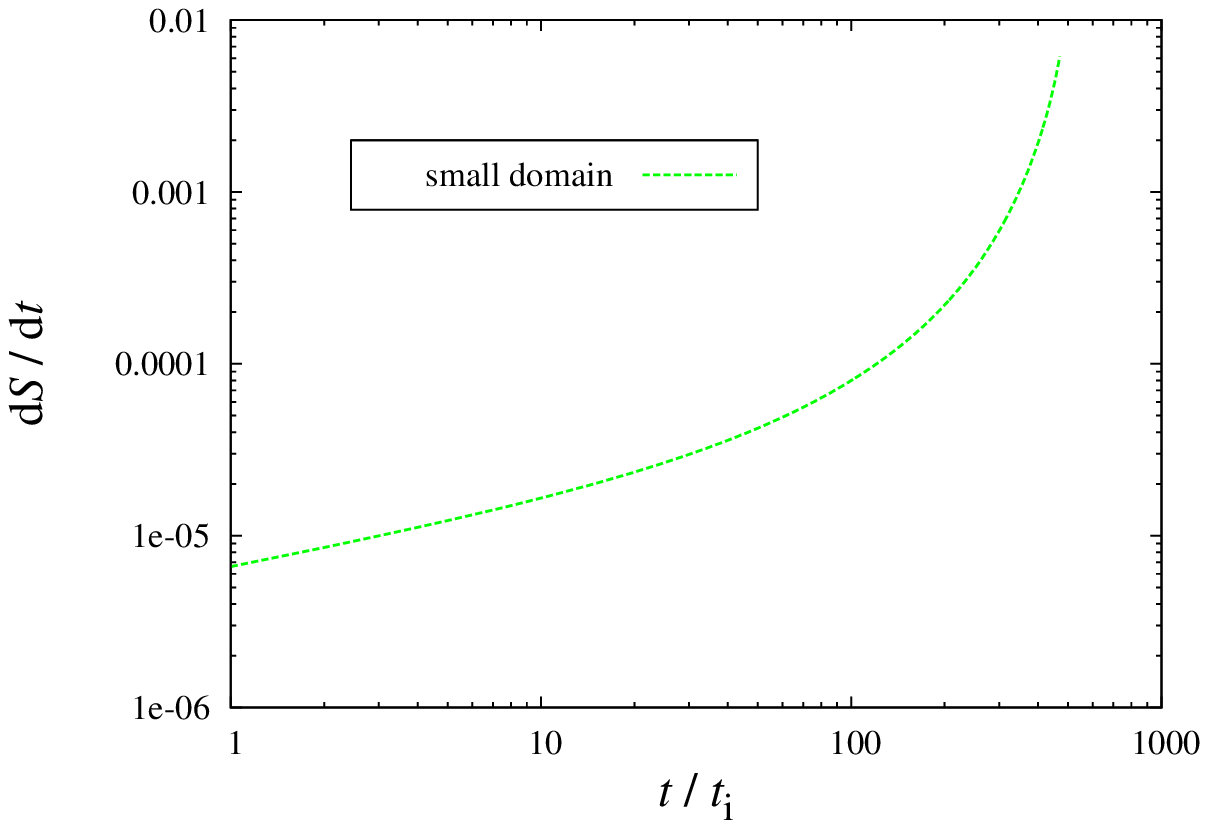}
  \caption{The same as Fig.~\ref{dotSlarge}, but for the small domain.} 
\end{figure}

Next we show the time evolution of the time derivative of the entropy, 
$\dot{\mathcal{S}}(t)$, in Figs.~\ref{dotSlarge} (for the large domain) 
and \ref{dotSsmall} (for the small domain). 
It should be emphasized that $\dot{\mathcal{S}}$ is positive 
in both cases throughout the evolution in our investigation. 
In the case of the large domain, however, the $\dot{\mathcal{S}}$ 
decreases with time temporarily, meaning that the second time derivative 
$\ddot{\mathcal{S}}$ is negative at early times. 
We presume that this result is caused by the vacuum region contained 
in the large domain, because the vacuum is a completely nonlinear structure 
where the linear perturbation theory does not apply, 
while the behavior of nonlinear clumps is somewhat similar to that of 
the linear perturbation at the early stage of the evolution. 
Detailed investigation is needed on this point. 

\section{Summary and Outlook}
\label{sec:sum}

In this paper, we study the time evolution of the Kullback-Leibler 
relative information entropy for cosmic matter distribution 
which was proposed as a measure of cosmic inhomogeneity 
in our previous work~\cite{hbm2004}. 
Intuitively this entropy is likely to possess the time-increasing nature, 
but it generally depends on initial conditions. 
We employ the linear perturbation of a spatially flat FLRW universe 
and the LTB model to demonstrate that the relative information entropy 
is convex and increases in time. 

We obtain explicit forms of the time derivative and 
the second time derivative of the entropy 
through the linear perturbation calculation. 
They tell us that in the linear regime of cosmic structure formation, 
the entropy is shown to be an increasing function of time at least 
 for sufficiently late times, 
even if the entropy decreases temporarily at early times. 
We also give an illustrative example of the time evolution of the entropy 
with a nonlinear LTB solution. 
This illustration also supports the time-increasing nature of the entropy 
although it shows that the entropy is not always time-convex. 
It seems from the example whether or not the measure is time-convex 
depends on the fraction of devoid regions within the averaging domain, 
i.e. on whether the universe is void-dominated. 
This will be further investigated in detail in a forthcoming publication.


\begin{theacknowledgments}
We would like to thank Jean-Michel Alimi for providing us 
an opportunity to present this work in the conference. 
We also thank Filipe C. Mena and Masahiro Morikawa for 
stimulating discussions and valuable remarks on gravitational entropy. 
The work of MM was partially supported by Grant-in-Aid for 
Young Scientists (B) (No.~18740164 and No.~21740199) from 
Japanese Ministry of Education, Culture, Sports, Science and Technology. 
This work is also supported by 
``F\'ed\'eration de Physique Andr\'e-Marie Amp\`ere, Lyon''.
\end{theacknowledgments}

\bibliographystyle{aipproc}   

\end{document}